\documentclass[twocolumn]{aastex63}
\usepackage{bm}

\newcommand{\mytilde}{\raise.19ex\hbox{$\scriptstyle\sim$}}

\received{}
\revised{}
\accepted{}

\submitjournal{}

\graphicspath{{./}}

\shorttitle{Free-Form SL Mass Reconstruction of the HFF clusters}
\shortauthors{Cha \& Jee}

\begin{document}

\title{Model-Independent Mass Reconstruction of the Hubble Frontier Field Clusters with MARS \\ 
Based on Self-Consistent Strong Lensing Data}

\correspondingauthor{M. James Jee}
\email{sang6199@yonsei.ac.kr, mkjee@yonsei.ac.kr}

\author{Sangjun Cha}
\affiliation{Department of Astronomy, Yonsei University, 50 Yonsei-ro, Seoul 03722, Korea}
\author{M. James Jee}
\affiliation{Department of Astronomy, Yonsei University, 50 Yonsei-ro, Seoul 03722, Korea}
\affiliation{Department of Physics and Astronomy, University of California, Davis, One Shields Avenue, Davis, CA 95616, USA}

\begin{abstract} 
We present new strong-lensing (SL) mass reconstruction of the six Hubble Frontier Fields (HFF) clusters with the MAximum-entropy ReconStruction ({\tt MARS}) algorithm. {\tt MARS} is a new free-form inversion method, which suppresses spurious small-scale fluctuations while achieving excellent convergence in positions of multiple images. For each HFF cluster, we obtain a model-independent mass distribution from the compilation of the self-consistent SL data in the literature. With $100-200$ multiple images per cluster, we reconstruct solutions with 
small scatters of multiple images in both source ($\mytilde0\farcs02$) and image planes ($0\farcs05-0\farcs1$), which are lower than the previous results by a factor of $5-10$.
An outstanding case is the MACS J0416.1-2403 mass reconstruction, which is based on the largest high-quality SL dataset where all 236 multiple images/knots have spectroscopic redshifts. 
Although our solution is smooth on a large scale, it reveals group/galaxy-scale peaks where the substructures are required by the data. We find that in general, these mass peaks are in excellent spatial agreement with the member galaxies, although {\tt MARS} never uses the galaxy distributions as priors.
Our study corroborates the flexibility and accuracy of the {\tt MARS} algorithm and demonstrates that {\tt MARS} is a powerful tool in the JWST era, when $2-3$ times larger number of multiple image candidates become available for SL mass reconstruction, and self-consistency within the dataset becomes a critical issue.  
\end{abstract}

\section{Introduction} 
\label{sec:intro}
Strong gravitational lensing (SL) is the most powerful method to reveal the detailed mass distribution in the central region of a galaxy cluster. A major strength of the SL analysis is that it can constrain the mass distribution at the cluster core with unparalleled S/N without any dynamical assumptions such as hydrostatic equilibrium \citep[e.g.,][]{kochanek2006, 2011A&ARv..19...47K}.
Although weak-lensing (WL) is useful in obtaining large-scale mass distributions, its precision is substantially inferior in the SL regime.

The application of SL science is not limited to its power to constrain the precise mass of a lens. We can probe the distant universe, utilizing the lensing magnification, which makes faint background sources brighter by a factor of tens to thousands
\citep[e.g.,][]{2022ApJ...931...81B, 2022Natur.603..815W, 2022arXiv221014123Y, 2022arXiv221015699W}.  
A study of time delays from multiply-imaged supernovae or quasars provides an independent measure of the Hubble constant $H_{0}$. 
Furthermore, the measurement of redshift-distance relations from the SL field with sources at different redshifts is a powerful probe of cosmological parameters including dark energy
\citep[e.g.,][]{2011MNRAS.411.1628D, 2018ApJ...865..122M, 2016ApJ...831..205K, 2016ApJ...817...60T, 2022ApJ...926...86A, 2022A&A...657A..83C, 2022arXiv221015794T}. 

The Hubble Frontier Fields (HFF) program  is designed to promote the SL-enabled sciences mentioned above with an emphasis on studying high-redshift galaxies through the magnification by massive gravitational lens telescopes
\citep{2015ApJ...800...84C,2017ApJ...837...97L}. 
The HFF sample consists of the six clusters:
Abell 370 ($z=0.375$, hereafter A370), Abell 2744 ($z=0.308$, hereafter A2744), Abell S1063 ($z=0.348$, hereafter AS1063), MACS J0416.1-2403 ($z=0.396$, hereafter MACSJ0416), MACS J0717.5-3745 ($z=0.545$, hereafter MACSJ0717), and MACS J1149.5-2223 ($z=0.543$, hereafter MACSJ1149). 
Thanks to the deep Hubble imaging data, hundreds of lensed multiple images have been identified from each galaxy cluster field and a number of studies have presented many interesting SL-related science results 
\citep[e.g.,][]{2014MNRAS.444..268R, 2015MNRAS.452.1437J, 2016ApJ...819..114K, 2016A&A...587A..80C, 2016MNRAS.459.3447D, 2016A&A...588A..99L, 2018MNRAS.473..663M, 2019MNRAS.485.3738L, 2021ApJ...919...54Z, 2021A&A...646A..57V, 2022arXiv220814020B, 2022arXiv220709416B}. 

Needless to say, in order to enable the aforementioned SL sciences, it is crucial to obtain a robust SL mass reconstruction. For example, a minor difference in mass distribution can lead to a considerable discrepancy in the prediction of magnification and time delay. Unfortunately, the literature has shown that there exist non-negligible differences among methods despite the use of hundreds of multiply-lensed images. This problem is best illustrated by the synthetic SL cluster analysis program \citep[][]{2017MNRAS.472.3177M}, where the mass-reconstruction participants are provided with identical sets of multiple images from two synthetic clusters whose true mass distributions remain hidden until the submission of the mass models by participants is complete.

Mathematically speaking, it is natural that the same multiple images, although there are hundreds, produce different results because the solution is still under-determined with respect to the mass resolution (i.e., degrees of freedom or complexity) that the SL-enabled science requires. In a sense, various mass reconstructions differ in how they overcome this ill-posed problem.

Another important issue in SL mass reconstruction is self-consistency within the employed dataset. Obviously, incorrect identification of even a few multiple image systems can have critical impacts on the resulting mass reconstruction. Although some systems have been robustly identified through spectroscopy or distinct morphology, a considerable number of multiple image systems were selected somewhat subjectively, based on weak evidence such as noisy photometric redshifts and intermediate mass models. If all multiple identifications are correct, an ideal solution should be able to predict all multiple image positions with negligible scatters. None of the existing SL algorithms has demonstrated such performances for the HFF sample yet. Also, few studies have suggested self-consistent SL datasets from the compilation of the literature catalogs.

Broadly, SL mass reconstruction methods are classified into two types: parametric and free-form approaches.
The parametric methods assume that the SL system is comprised of analytic halos such as elliptical/circular Navarro-Frenk-White profiles \citep[NFW;][]{1996ApJ...462..563N} that are spatially correlated with the cluster galaxies; because of the use of the light-mass correlation, the technique is also often referred to as a light-trace-mass (LTM) method.
They determine the optimal mass distributions by adjusting the halo properties in such a way that the resulting model reproduces the observed SL data as close as possible.
The free-form methods assume neither a particular analytic profile nor correlation between galaxies and dark matter, often reconstructing mass distributions on a grid.

In the case of the parametric methods, they have a relatively small number of free parameters, which makes the optimization relatively easy and fast. The weakness of this approach includes the lack of flexibility and model dependence. This often leads to less than optimal convergence of the multiple image positions in the source plane. Because of the limited degrees of freedom, the methods are not optimal if one is interested in testing unconventional cases, where the dark matter distribution deviates from the galaxy distribution. This includes probes of substructures devoid of luminous galaxies \citep[e.g., ][]{2011MNRAS.417..333M, 2012ApJ...747...96J, 2014ApJ...783...78J} or genuine galaxy-mass offsets due to self-interacting dark matter \citep[e.g.,][]{2008ApJ...679.1173R}.

The free-form methods require a relatively large number of free parameters. The superb flexibility allows the mass model to converge multiple image positions in the source/image plane with negligible scatter {\it if} they are real. Therefore, the methods can be used to test the self-consistency of the SL dataset up to the  flexibility of the algorithm.
The obvious drawback of this approach is degeneracy, which in general produces multiple solutions for the given set of SL data.
Therefore, some free-form methods adopt an average of $10-100$ solutions as their representative model.

The MAximum-entropy ReconStruction ({\tt MARS}) algorithm \citep{2022ApJ...931..127C} is a new grid-based free-form SL reconstruction method, which overcomes the aforementioned degeneracy problem without compromising the flexibility.
The {\tt MARS} approach adopts the maximum cross-entropy method \citep{1998MNRAS.299..895B, 2007ApJ...661..728J} as a regularization, which makes the solution quasi-unique and suppresses spurious substructures that arise in many free-form SL reconstructions.
The performance of {\tt MARS} was tested
with the public synthetic SL data of \cite{2017MNRAS.472.3177M}. The result shows that the mass reconstruction quality either exceeds or is on par with those of the best-performing methods \citep{2022ApJ...931..127C}. 

Having verified the high fidelity of the {\tt MARS} algorithm, in this study, 
we apply {\tt MARS} to the six HFF clusters.
Although a number of studies have presented SL studies of the six HFF clusters, the results vary quite significantly among different methods. The diversity is not surprising because, as mentioned above, even hundreds of multiple images cannot break the degeneracy. Also, the authors do not use the same SL dataset, as new multiple images keep becoming available.
The current study is based on our compilation of $100 - 200$ self-consistent multiple images per cluster drawn from the literature. Since we find that {\tt MARS} is one of the best-performing methods when tested with the public synthetic SL data, 
we believe that our mass reconstruction results are useful contributions to the SL community. Therefore, we make our results and compiled catalogs publicly available along with the code sets, with which interested readers can verify our results in this paper (\dataset[doi:10.5281/zenodo.7804575]{https://doi.org/10.5281/zenodo.7804575}).

This paper is organized as follows. In \textsection\ref{sec:method}, we review our {\tt MARS} algorithm briefly. 
\textsection\ref{sec:data} describes the compilation of the multiple images and the publicly available SL models.
\textsection\ref{sec:result} presents our mass reconstruction results.
In \textsection\ref{sec:comparison}, we compare our results with the previous ones before we conclude in \textsection\ref{sec:conclusion}. 
Unless stated otherwise, this paper assumes a flat $\Lambda$CDM cosmology with the matter density parameter $\Omega_{m}=0.27$ and the dimensionless Hubble constant parameter $h=0.72$.

\section{Method} \label{sec:method}
{\tt MARS} is a free-form SL reconstruction method based on source plane minimization and maximum-entropy regularization.
Here, we summarize the main 
algorithms
of the {\tt MARS}. For 
details, we refer readers to 
\citet{2022ApJ...931..127C}.

The relation between the image position $\bm{\theta}$ and the source position $\bm{\beta}$ follows the non-linear lens equation:
\begin{equation}
    \bm{\beta}=\bm{\theta}-\bm{\alpha}(\bm{\theta}),
\label{lens_equation}
\end{equation}
where $\bm{\alpha}$ is the deflection angle. 
The deflection angle $\bm{\alpha}$ can be computed through the derivative of the deflection potential $\Psi$ or the convolution of the convergence $\kappa$. 
{\tt MARS} uses the convolution to calculate the deflection angle $\bm{\alpha}$:
\begin{equation}
    \bm{\alpha} (\bm{\theta}) = \frac{1}{\pi} \int
    \kappa (\bm{\theta}^{\prime}) \frac{\bm{\theta}-\bm{\theta}^{\prime}}{|\bm{\theta}-\bm{\theta}^{\prime}|^{2}} \bm{d^{2} {\theta}}^{\prime}. 
    \label{eqn_deflection_via_con}
\end{equation}
The convergence $\kappa$ is the unitless surface mass density defined as:
\begin{equation}
    \kappa=\frac{\Sigma}{\Sigma_{c}},
\label{eqn_kappa}
\end{equation}
where $\Sigma$ ($\Sigma_c$) is the (critical) surface mass density.
The $\Sigma_c$ is defined by the following:
\begin{equation}
    \Sigma_{c}=\frac{{c^2}D_s}{4{\pi}G{D_{ds}}{D_d}},
\end{equation}
where $D_{ds}$ is the angular diameter distance between the source and lens, $D_s$ is to the source, and $D_d$ is to the lens.

{\tt MARS} converges multiple image positions in the source plane. That is,
we minimize the scatters of the delensed positions of the multiple images.
Our $\chi^2$ function to minimize is:
\begin{equation}
    \chi^{2}=\sum_{i=1}^{I} \sum_{j=1}^{J}\frac{(\bm{\theta}_{i,j}-\bm{\alpha}_{i,j}(z)-\bm{\beta}_{i})^{2}}{{\sigma_{i}}^{2}},
\label{eqn_chi_squared}
\end{equation}
where
\begin{equation}
    \bm{\beta}_{i}=\frac{1}{J}\sum_{j=1}^{J}(\bm{\theta}_{i,j}-\bm{\alpha}_{i,j}(z)).
\end{equation}
$I$ is the total number of sources and $J$ is the number of multiple images from each source. We considered each identified knot across its multiple images a separate image.

Caution is needed when computing scatters because a solution favoring a higher magnification can provide a smaller scatter. Therefore, when constructing our $\chi^2$ function, we let the uncertainties vary according to the magnification.
This is implemented by creating two virtual knots for each source position in the image plane. They are offset from the source center in mutually perpendicular directions.
Uncertainties are evaluated by first computing the locations of the delensed knots (including both multiple images and virtual knots) in the source plane and then finding the maximum distances between knots along the $x$- and $y$-axes (see Figure 1 in \citet{2022ApJ...931..127C}).
In \citet{2022ApJ...931..127C}, the offset was set to $0\farcs2$ for all sources, which led to a mean image plane scatter of $\mytilde0\farcs01$. Since this scatter is five times smaller than the native pixel size ($\mytilde0\farcs05$) of the Hubble Space Telescope (HST)/Advanced Camera for Surveys (ACS), slight overfitting occurred.
Thus, in this study, we start our mass reconstruction with a virtual knot distance of $1\arcsec$ for all sources. This makes the mean image plane scatters increase to $0\farcs05-0\farcs1$.

Not all multiple images have well-defined centroids. Some multiple images were identified only in the MUSE data. Also, quite a few multiple images have a diffuse surface brightness distribution.
For these large-centroid-uncertainty (LCU) images,
We increase the virtual knot distances 
to $5\arcsec$, which
relaxes the convergence constraints by a factor of 5.
In the case of MACSJ1149, 
we increase the virtual knot distances to $2\arcsec$ for all multiple images because most of the multiple images are located in the high surface brightness region affected by the intracluster light or bright galaxies, and their centroid uncertainties are relatively large.

The regularization term is important to prevent overfitting and achieve smooth mass maps. 
{\tt MARS} regularizes the $\chi^{2}$ minimization process by maximizing the following cross-entropy $R$:
\begin{equation}
    R=\sum\left (p-\kappa+\kappa\mathrm{ln} \frac{\kappa}{p} \right),
\label{cross_entropy}
\end{equation}
where $p$ and $\kappa$ are the prior and current convergence, respectively. The prior is the Gaussian-smoothed version of the convergence map that is obtained from the previous minimization run.

Finally, {\tt MARS} minimizes the following function:
\begin{equation}
    f={\chi}^{2}+rR, 
\label{total_equation}
\end{equation}
where $r$ is the regularization control parameter.
Our mass reconstruction is performed in two steps. 
First, we reconstruct a low-resolution mass distribution with a $70\times70$ mass grid (the actual reconstruction field is $50\times50$ because the outer 10-element thick boundary is used as a margin). We use a flat $\kappa=0.5$ map for the initial prior. As mentioned in \cite{2022ApJ...931..127C}, the {\tt MARS} mass reconstruction is not sensitive to the choice of the initial prior.
After this low-resolution mass reconstruction is obtained, we increase the resolution by a factor of two ($140\times140$ grid including the margin) and reconstruct a high-resolution mass map. We use the low-resolution result as the initial prior for this second step.
Therefore, the number of free parameters
needed to represent the final mass map is extremely large ($140\times140=19600$).

In addition to the parameters for the mass map, we treat the redshifts of some multiple images as free parameters.
These are the multiple images that have only a model redshift\footnote{The model redshift means the redshift value favored by the given mass model.} or cannot be converged using the photometric redshift reported in the literature.
We set a flat prior for a model redshift as $z_{model} = [z_{cluster} + 0.1 , 15]$, where $z_{cluster}$ is the redshift of the galaxy cluster.

\section{Data} \label{sec:data}

\subsection{Multiple Images for SL Modeling} \label{sec:multiple_data}
\begin{figure*}
\centering
\includegraphics[width=0.87\textwidth]{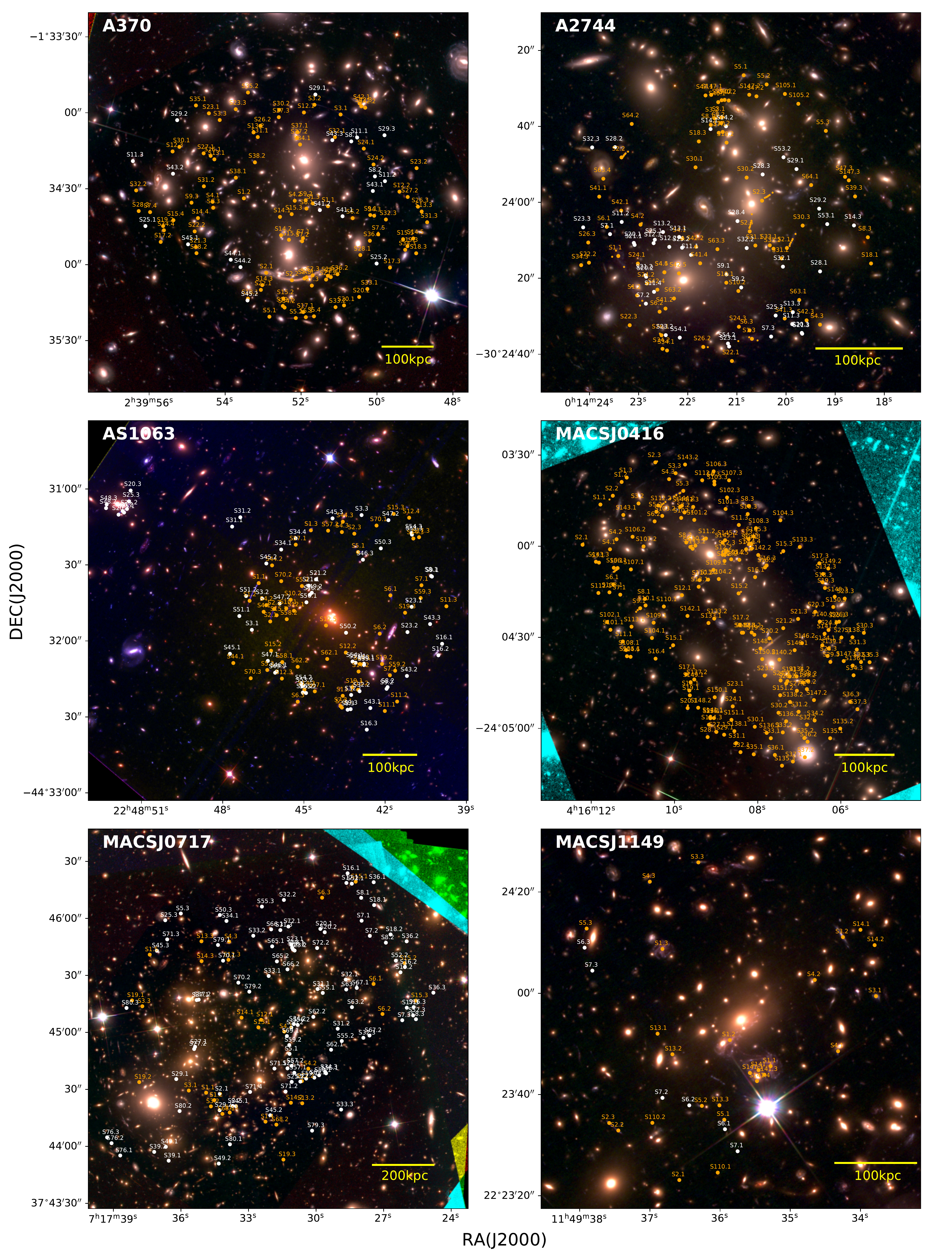} 
\caption{Multiple image distributions in the six HFF cluster fields. Circles and numbers in orange (white) indicate the locations and IDs of the gold (silver) multiple images, respectively. Orange stars represent the knots of the extended multiple images that are also utilized for reconstruction. The color-composite images are created with the F606W, F814W, and F105W filters 
to represent intensity in blue, green, and red, respectively.
The exception is AS1063, for which we use F435W, F606W, and F814W for the blue, green, and red channels, respectively.}
\label{multiple_catalog_fig}
\end{figure*}

From the literature, we compiled 
multiple images in the six HFF cluster fields.   
We classified multiple images into two classes: gold and silver.
The gold class images have spectroscopic redshifts and thus are considered more
secure. 
The silver class includes the sources that have only photometric redshifts. The sources that do not have photometric redshifts, but are agreed to be bona-fide multiply-lensed images by multiple studies are also classified as the silver class.
Their redshifts are treated as free parameters and constrained together with the mass reconstruction (i.e., model redshift).
For the silver class images with photometric redshifts, we adopted the reported photometric redshifts as their redshifts if they converge in the source plane along with the rest of the images. In case they do not converge with the reported redshifts, we also regarded their redshifts as free parameters and
determined their model redshifts. Although it is rare (2 cases for MACS0717 and MACS0416), we do not use the sources that {\tt MARS} cannot make converge (see the discussion for these two cluster results).
When a source possesses distinct knots that can be identified consistently across
its multiple images, we handled each knot as if it is a separate multiple image.
Note that the total number of multiple images that we quote hereafter is computed by counting knots as different multiple images. 

Figure~\ref{multiple_catalog_fig} shows the multiple image distributions in the six HFF clusters, which we use for mass reconstruction in this paper.
Each cluster has $100-200$ multiple images, all of which are self-consistent in the sense that {\tt MARS} can converge them simultaneously in both source and image planes.
Our multiple image tables are
provided in Appendix~\ref{multiple_image_table}. 
In Table~\ref{public_SL_table}, we summarize the plate scale, critical surface mass density, and the number of multiple images for the six HFF clusters.
Below we provide details for individual clusters. 
We listed multiple images for SL modeling in tables which are available as online supplementary materials.

\subsubsection{Abell 370}
We combined the multiple image catalogs reported by \citet{2019MNRAS.485.3738L} and \citet{2021MNRAS.506.6144G} for A370. 
The spectroscopic redshifts in their catalog are observed by the Multi-Unit Spectroscopic Explorer Guaranteed Time Observation (MUSE GTO) program \citep[]{2017MNRAS.469.3946L, 2019MNRAS.485.3738L}. We selected 119 multiple images (100 gold and 19 silver images) to reconstruct the SL model of A370. Among the silver images, the redshifts of sources 8, 11, 41, 44, and 45 are free parameters and determined by {\tt MARS}.
In A370, the number of the LCU images is 21, for which we relaxed their convergence constraints as mentioned in \textsection\ref{sec:method}.
In the upper left panel of Figure~\ref{multiple_catalog_fig}, we show the multiple image distribution in the central $150\arcsec\times150\arcsec$ region, for which we perform mass reconstruction.
Our A370 multiple images are listed in Table~\ref{multiple_catalog_a370}.

\subsubsection{Abell 2744}\label{data_a2744}
We used the catalogs of \citet{2015MNRAS.452.1437J}, \citet{2016ApJ...819..114K}, \citet{2018MNRAS.473..663M}, and \citet{2022arXiv220709416B} for A2744. Among the multiple images without a spectroscopic redshift, the images agreed by the first three papers are classified as silver; all multiple images in \citet{2022arXiv220709416B} have spectroscopic redshifts. We adopt the model redshifts derived by \citet{2018MNRAS.473..663M} for silver images. The knots from sources 1, 2, 3, 4, and 26 identified by \citet{2022arXiv220709416B} are included.
We excluded images 33.3, sources 61, and 62, which are identified with the MUSE data, but not detected in the deep JWST image \citep[e.g.,][]{2022arXiv221204026B, 2023arXiv230102671W}.
A total of 162 images (120 gold and 42 silver images), which include 43 knots, are used for the final mass reconstruction.
In the upper right panel of Figure~\ref{multiple_catalog_fig}, we show the multiple image distribution in the central $100\arcsec\times100\arcsec$ region, for which we present SL mass reconstruction.
Table 3 lists our Abell~2744 multiple images.

\subsubsection{Abell S1063}
The multiple images from \citet{2014MNRAS.444..268R}, \citet{2016A&A...587A..80C}, \citet{2016MNRAS.459.3447D}, and \citet{2018ApJ...855....4K} are combined for AS1063. 
The multiple images that are used by at least three papers are
classified as silver.
We determined the model redshifts for all silver images for this system.
The total number of multiple images is 116 (57 gold and 59 silver images). We identified 3 images as LCU objects.
We display the $150\arcsec\times150\arcsec$ color-composition image of Abell S1063 in the middle left panel of Figure~\ref{multiple_catalog_fig}. 
We compiled the SL catalog in Table 4, following the numbering scheme of \citet{2014MNRAS.444..268R}.

\subsubsection{MACS J0416.1-2403}
We combined the catalogs of \citet{2021A&A...646A..57V} and \citet{2021A&A...646A..83R} for MACSJ0416.
All multiple images in this cluster have spectroscopic redshifts from the two MUSE programs: 094.A-0115B (northeast region, PI: J. Richard and 0100.A-0763A) and 
094.A-0525A (southwest region, PI: F.E. Bauer).
The two papers have different numbering schemes and we based our compilation on the \citet{2021A&A...646A..57V} scheme. We find that the multiple image 55.2 in \citet{2021A&A...646A..83R}  \citep[system 11 in][]{2021A&A...646A..57V} cannot be converged by {\tt MARS}. Also, the mass reconstruction with the inclusion of it produces ``pinched" features.
Since our test with the mock SL data
\citep{2022ApJ...931..127C} shows that
{\tt MARS} can converge all multiple images without such artifacts if they are bona fide multiple images, the creation of this unphysical substructure indicates that perhaps this multiple image is not real.
Thus, we excluded this image for reconstruction.
From visual inspection, we identified 32 images as LCU objects and relaxed their convergence constraints (\textsection\ref{sec:method}).
The total number of multiple images including knots in the compiled catalog is 236. Since all sources in MACSJ0416 have spectroscopic redshifts. In terms of both quantity and quality, the dataset of the cluster is regarded as a precursor to the JWST SL data.
In this study, we reconstruct the central $125\arcsec\times125\arcsec$ region of the cluster shown in the middle right panel of Figure~\ref{multiple_catalog_fig}. 
We list the multiple images in Table 5.

\subsubsection{MACS J0717.5+3745}
For MACSJ0717, we use the multiple images identified by \citet{2016A&A...588A..99L}. We find that the multiple images from source 75 cannot be converged even though we let {\tt MARS} treat its redshift as a free parameter (source 75 has only a model redshift). Given the flexibility of {\tt MARS}, this indicates that perhaps the multiple images from source 75 may not comprise a true lensed system. We did not use source 75 in our final mass reconstruction.
The total number of multiple images for MACSJ0717 is 131 (31 gold and 100 silver images, Table 6).
Only two images are identified as LCU objects.
The lower-left panel of Figure~\ref{multiple_catalog_fig} displays the central 
$200\arcsec\times200\arcsec$ region of the cluster, for which we reconstruct the mass map.

\subsubsection{MACS~J1149.5+2223}
To reconstruct the mass model of MACSJ1149, we adopt the multiple images of \citet{2021ApJ...919...54Z}, which include 94 knots from the lensed host galaxy of
SN Refsdal. 
We set the virtual knot distances to $2\arcsec$ for all multiple images because most of them are located in the high surface brightness region affected by the intracluster light or cluster members, and their centroids are more uncertain.
Our mass reconstruction is limited to the central $75\arcsec\times75\arcsec$ region, which is displayed in the lower right panel of
Figure~\ref{multiple_catalog_fig}.

\subsection{Public SL Models in the literature}\label{data_public_SL}
We compare our mass reconstruction results with those in the public domain\footnote{\url{https://archive.stsci.edu/prepds/frontier/lensmodels/}}.
There are multiple mass models from the same team/method and we use the constraints from version 4. We list the compared models for each HFF cluster in Table~\ref{public_SL_table}.

\begin{deluxetable*}{ccccccc}\label{public_SL_table}
\tablecaption{Summary of the six HFF cluster data.}
\tablehead {
\colhead{Cluster} &
\colhead{Redshift} &
\colhead{Plate Scale} &
\colhead{Critical Surface Density$^1$} & 
\colhead{Gold (knot)$^2$} &
\colhead{Silver$^3$} &
\colhead{Compared Models$^4$} \\
\colhead{} &
\colhead{} &
\colhead{(kpc/$\arcsec$)} &
\colhead{($10^{9}M_{\odot}/~\rm kpc^{2}$)} & 
\colhead{} &
\colhead{} &
\colhead{}
}
\startdata
A370 &  0.375 & 5.060 & 1.593 & 100 (0) & 19 & D, C, G, S/J, K, Gr, B \\
A2744 &  0.308 & 4.439 & 1.816 & 120 (43) & 42 & D, C, G, S/J, K, Gr \\
AS1063 &  0.348 & 4.821 & 1.672 & 57 (0) & 59 & D, C, G, S/J, K, Gr \\
MACSJ0416 &  0.396 & 5.236 & 1.540 & 236 (46) & 0 & D, C, G, S/J, K, Gr \\
MACSJ0717 &  0.545 & 6.276 & 1.285 & 31 (0) & 100 & D, C, S/J, K, Gr \\
MACSJ1149 &  0.543 & 6.264 & 1.287 & 117 (94) & 6 & D, C, S/J, K, Gr \\
\enddata
\tablecomments{$^1$Evaluated with $D_{ls}/D_s=1$. $^2$Number of gold images, which includes the number of knots. $^3$Number of silver images.   $^4$For comparison, we use the lens models from Diego (D), CATS (C), GLAFIC (G), Sharon/Johnson (S/J), Keeton (K), GRALE (Gr), and Brada{\v{c}} (B).}
\end{deluxetable*}

We select both parametric and free-form SL algorithms. For the parametric method, we use CATS \citep[]{2009MNRAS.395.1319J,2012MNRAS.426.3369J,2014MNRAS.443.1549J,2014MNRAS.444..268R}, GLAFIC \citep[]{2010PASJ...62.1017O,2015ApJ...799...12I,2016ApJ...819..114K,2018ApJ...855....4K}, Sharon/Johnson \citep[]{2014ApJ...797...48J}, and Keeton \citep[]{2010GReGr..42.2151K,2014ApJ...781....2A,2014MNRAS.443.3631M}.
For the free-form method, we choose Diego\footnote{Some classify that the SL method of Diego, WSLAP+ is a hybrid method because WSLAP+ includes compact mass components to describe cluster members \citep{2017MNRAS.472.3177M}.} \citep[]{2015MNRAS.447.3130D,2015MNRAS.451.3920D,2016MNRAS.456..356D,2016MNRAS.459.3447D,2018MNRAS.473.4279D}, {\tt GRALE} \citep[]{2006MNRAS.367.1209L,2016MNRAS.461.2126S}, and Brada{\v{c}} \citep[]{2005A&A...437...39B}.

Because the public $\kappa$ maps are scaled to $D_{ds}/D_{s}=1$, we also rescale our mass models using the same scheme for a fair comparison. In the case of the magnification ($\mu$), the results in this paper present magnifications for a source at $z = 9$.
Readers are reminded that our multiple-image catalogs are not identical to those used by these models in the literature.

\section{Result} \label{sec:result}

\subsection{Image Plane Scatters of the Multiple Images}\label{sec:SL_model_best}
\begin{figure*}
\centering
\includegraphics[width=0.98\textwidth]{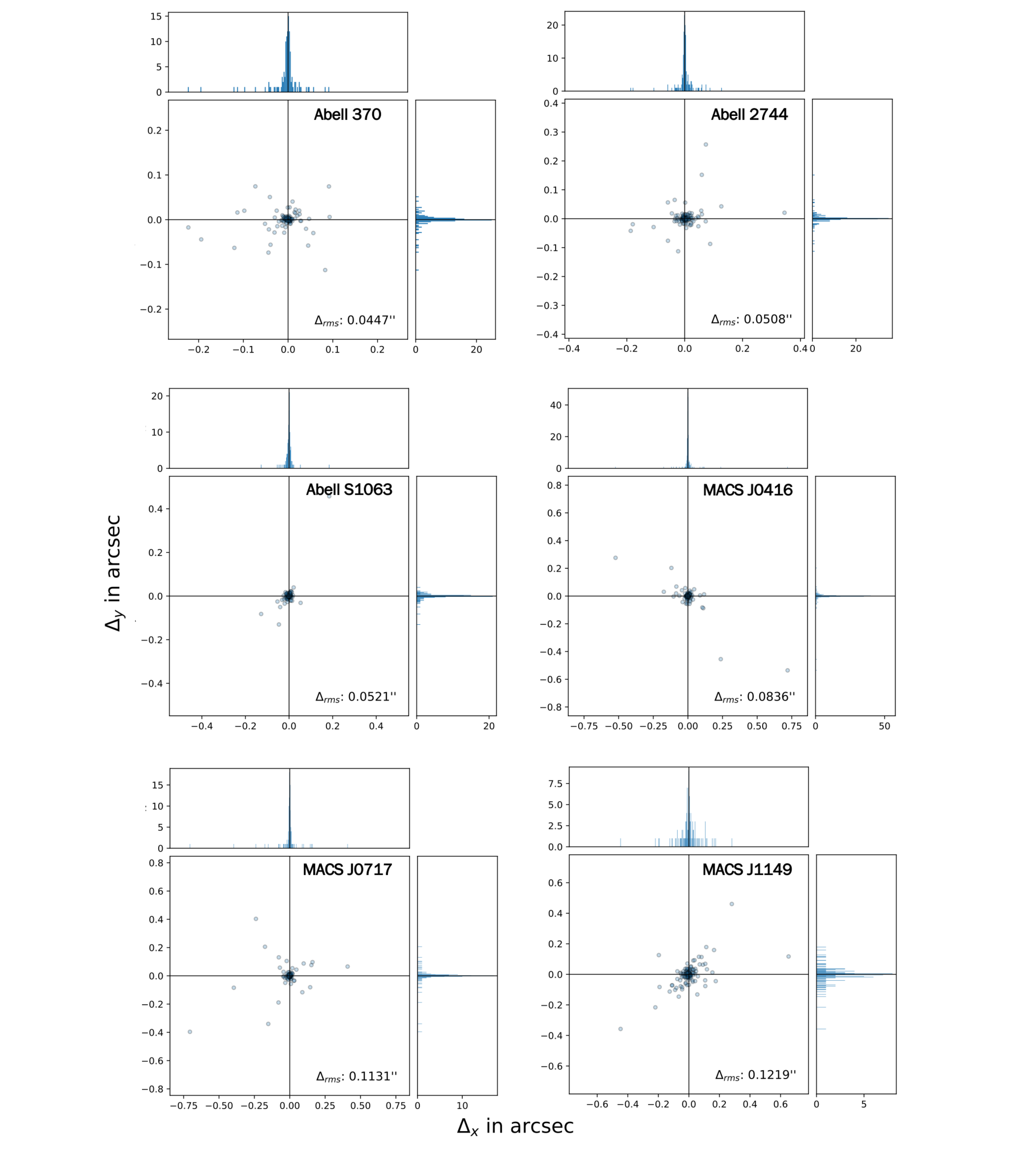} 
\caption{Scatters between the observed and predicted multiple images in the image plane. For each source, we used the barycenter in the source plane to find the multiple image positions in the image plane.$\Delta_{x}$ ($\Delta_{y}$) indicates the displacements from the observed multiple images along the $x$-axis ($y$-axis). $\Delta_{rms}$ shows the total rms value of the image plane scatters (see Equation~\ref{image_scatter})}. 
\label{lens_scatter}
\end{figure*}

\begin{figure}
    \plotone{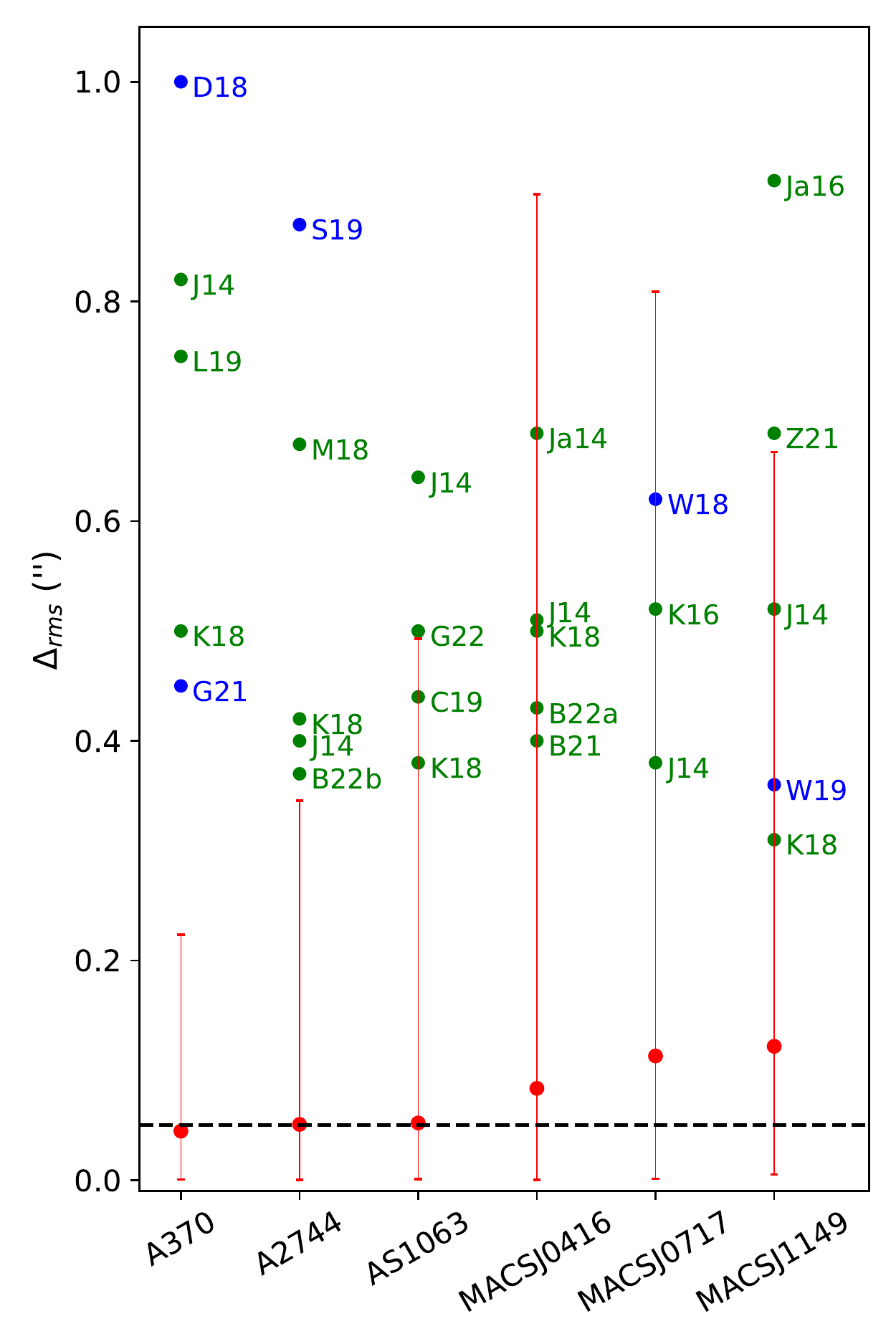}
    \caption{Image plane scatter comparison between {\tt MARS} and the literature results. The green (blue) dots show the rms from the parametric (free-form) methods. The red dots display the result from {\tt MARS}.
    The error bars represent the entire scatter range (between the minimum and maximum), not the 68\% interval. The black dashed line indicates an rms value of $0\farcs05$, which is the pixel size of the ACS detector. We adopt the lens plane rms values from \citet[J14]{2014ApJ...797...48J}, \citet[Ja14]{2014MNRAS.443.1549J}, \citet[Ja16]{2016MNRAS.457.2029J}, \citet[K16]{2016ApJ...819..114K}, \citet[D18]{2018MNRAS.473.4279D}, \citet[K18]{2018ApJ...855....4K}, \citet[M18]{2018MNRAS.473..663M}, \citet[W18]{2018MNRAS.480.3140W}, \citet[C19]{2019A&A...632A..36C}, \citet[L19]{2019MNRAS.485.3738L}, \citet[S19]{2019MNRAS.488.3251S}, \citet[W19]{2019MNRAS.482.5666W}, \citet[B21]{2021A&A...645A.140B}, \citet[G21]{2021MNRAS.506.6144G}, \citet[Z21]{2021ApJ...919...54Z}, \citet[B22a]{2022arXiv220814020B}, \citet[B22b]{2022arXiv220709416B}, and \citet[G22]{2022A&A...659A..24G}.} 
    \label{fig:img_scatter_comparison}
\end{figure}

In Figure~\ref{lens_scatter}, we plot two-dimensional multiple-image scatter distributions in the image plane. 
To quantify the performance of each mass model in terms of image position accuracy, we calculate the root-mean-square (rms) of the scatters between predicted and observed multiple image positions in the image plane as follows:
\begin{equation}\label{image_scatter}
    \Delta_{rms}=\sqrt{\frac{1}{M}\sum_{m=1}^{M}|\bm{\theta}_{truth,m}-\bm{\theta}_{model,m}|^{2}},
\label{eqn_rms}
\end{equation}
where $M$ is the total number of multiple images, and $\bm{\theta}_{truth,m}$ and $\bm{\theta}_{model,m}$ are the locations of the observed and predicted multiple images for image $m$, respectively.

For the entire sample, our results produce small scatters in both the source plane ($\mytilde0\farcs02$) and the image plane ($0\farcs05-\farcs1$). Readers are reminded that we allowed some LCU images to have larger image-plane scatters to prevent possible overfitting
(\textsection\ref{sec:method}). Our average image-plane scatters are a factor of $5-10$ smaller than those in the previous studies as shown in Figure~\ref{fig:img_scatter_comparison}.
Interestingly, the scatters from the previous free-form methods (e.g., D18, W18, S19, W19, and G21) are not smaller than those from the parametric models, often exceeding $\sim0\farcs4$. 
The most remarkable case is MACSJ0416. We achieved $\Delta_{rms}=0\farcs084$ even with the largest number (236) of multiple images, all of which have spectroscopic redshifts.

Our mass models provide the smallest image-plane scatters, and this is clearly one of the strengths of {\tt MARS}.
Nevertheless, it is important to note that the scatter size is only one of many metrics to evaluate the quality of the lens model. That is, a reasonably good convergence in the image/source plane is a necessary condition for a robust lens model close to the truth, although it is not a sufficient condition. Another useful metric to assess the model fidelity is a lens-plane image reconstruction (relensing).
Both the morphology and surface brightness features predicted by a robust model are expected to closely follow the observation.
We present our lens-plane image reconstruction results in
Appendix~\ref{lens_plane_recon}.

\subsection{Mass Model}\label{sec:mass_model}
\begin{figure*}
\centering
\includegraphics[width=0.88\textwidth]{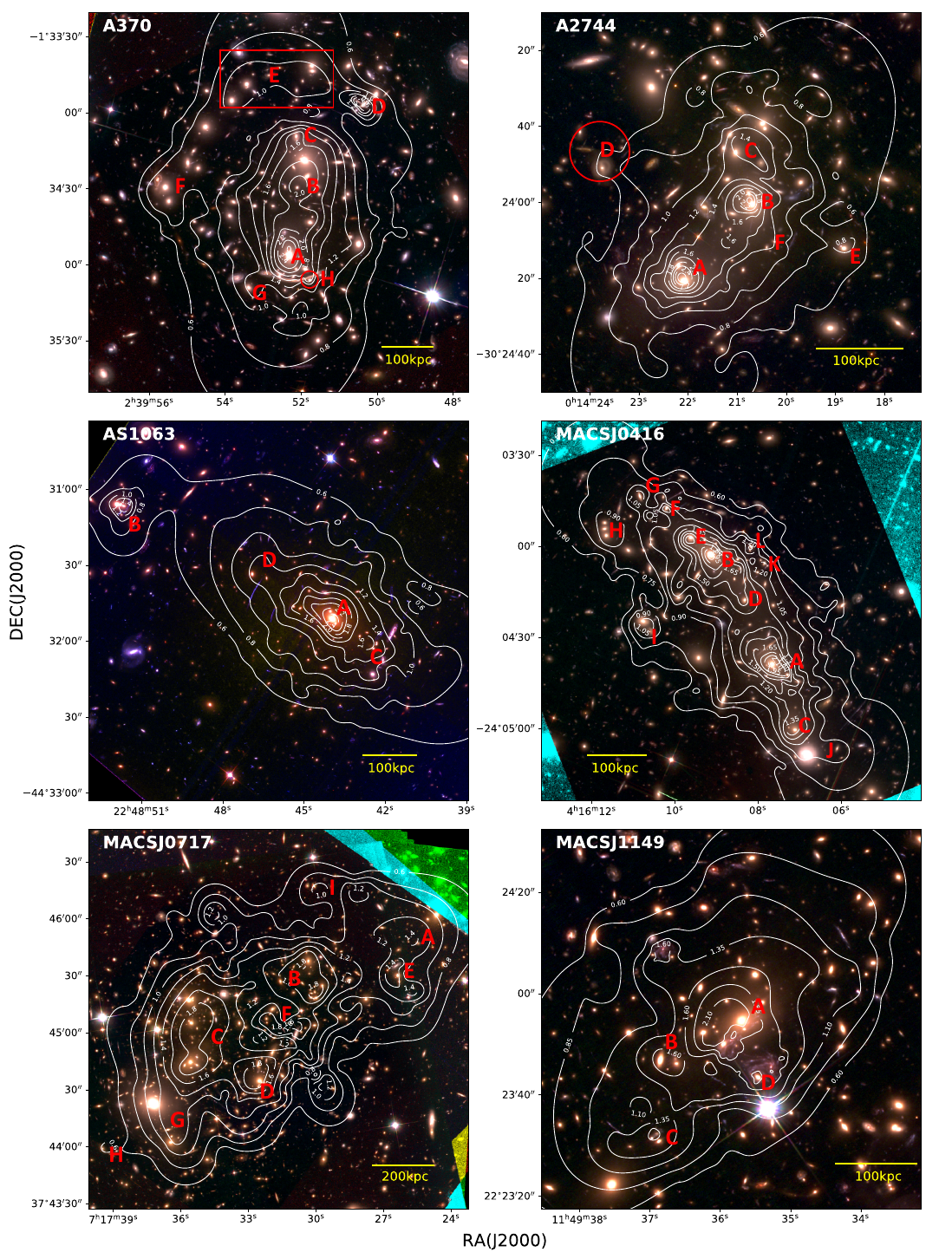} 
\caption{Mass contours of the HFF clusters overlaid on the composite color images. The white contours indicate the convergence $\kappa$. The red labels mark the locations of the substructures that are cospatial with the spectroscopically confirmed cluster members.
The color-composite images are the same as in Figure~\ref{multiple_catalog_fig}.}
\label{result_kappa}
\end{figure*}

We present the reconstructed $\kappa$ contours of the six HFF clusters in Figure~\ref{result_kappa}. 
Below we discuss the result for each cluster.

{\bf A370.} The overall mass distribution is characterized by the two large mass clumps
separated by $\mytilde150$~kpc.
The stronger of the two is the southern halo (A), which coincides with the southern BCG. The northern mass peak (B) is $\mytilde50$~kpc offset toward the south from the northern BCG. 
A similar offset is present in both free-form \citep[e.g.,][]{2018MNRAS.473.4279D,2021MNRAS.506.6144G} and parametric mass reconstructions \citep[e.g.,][]{2014ApJ...797...48J, 2018ApJ...855....4K, 2019MNRAS.485.3738L}.
In addition to these two main halos, our {\tt MARS} mass reconstruction also reveals several group/galaxy-scale substructures at the locations of the bright cluster galaxies (C, D, F, G, and H); the substructures C, D, and H are referred to as G2, G4, and G1 in \citet{2019MNRAS.485.3738L}, respectively. This is remarkable because previous free-from methods did not detect these substructures as isolated peaks. The most compact substructure among these is the substructure H (centered at the encircled galaxy), which is constrained by the neighboring SL features including the giant ``cosmic snake."
One peculiar substructure in our mass map is the east-west-elongated overdense region (E), which is $\mytilde40\arcsec$ north of the northern BCG. Our overlay shows that this diffuse substructure is traced by some cluster member galaxies.

{\bf A2744.} Similar to A370, the cluster mass distribution is dominated by two main halos (A and B), which coincide with the two brightest cluster members. The distance between the two mass clumps is $\mytilde100$~kpc.
The third most significant mass clump in A2744 is clump C, which is $\mytilde80$~kpc north of the northern BCG. This substructure is also reported by some free-from \citep[e.g.,][]{2015ApJ...811...29W,2019MNRAS.488.3251S} and many parametric approaches \citep[e.g.,][]{2016MNRAS.463.3876J, 2018MNRAS.473..663M, 2022arXiv220709416B}. Our {\tt MARS} mass reconstruction detects some group/galaxy-scale substructures (D, E, and F) 
at the location of the bright cluster members.

{\bf AS1063.} The mass model of AS1063 consists of the main halo (A) centered at the BCG and the sub-halo (B) centered at the northeastern cluster member. The distance between the two mass clumps is $\mytilde450$~kpc. 
This northeast-southwest elongation was also reported in the literature \citep[e.g.,][]{2014MNRAS.438.1417M, 2014ApJ...797...48J, 2016A&A...587A..80C, 2016MNRAS.459.3447D, 2018ApJ...855....4K, 2022A&A...659A..24G}.
Along the axis defined by A and B are substructures C and D, which coincide with the cluster members.

{\bf MACSJ0416.} The cluster has the largest (236) number of multiple images among the six HFF clusters. Furthermore, all multiple images have spectroscopic redshifts. This exceptionally high-quality SL data enables {\tt MARS} to achieve a high-fidelity, high-resolution mass reconstruction. The small image-plane scatters ($\mytilde0\farcs044$) indicate that all 236 multiple images are self-consistent and perhaps real. We emphasize that none of the SL studies in the literature achieved this level of precision \citep[e.g.,][]{2018ApJ...863...60R, 2021A&A...645A.140B, 2022arXiv220814020B}. 

The overall mass distribution is elongated in the northeast-southwest orientation. The
two strongest mass peaks are A and B, which are centered at the two BCGs separated by $\mytilde200$~kpc.
In addition to these two main mass peaks, {\tt MARS} detects a number of
group/galaxy-scale mass peaks (C through L), all of which are in good spatial agreement with
the cluster galaxies. This high-resolution mass reconstruction is unprecedented for the free-form method \citep[e.g.,][]{2015MNRAS.447.3130D, 2016MNRAS.461.2126S}.

{\bf MACSJ0717.} The overall cluster mass distribution shows a northwest-southeast elongation. The reconstructed mass map is complex and reveals a number of substructures. However, unlike in the case of MACSJ0416, not all substructures possess clear corresponding cluster members. The cluster has the smallest number (33) of gold-class multiple images in the HFF sample, which are significantly outnumbered by the silver-class images (100). MACSJ0717 is one of the prime targets in the HFF sample, whose understanding of the mass distribution needs to be improved by future spectroscopic observations. 

Our reconstruction reveals clumps A, B, C, D, and E, which are consistent with previous results except for clump A \citep[e.g.,][]{2011MNRAS.410.1939Z, 2015MNRAS.451.3920D, 2016A&A...588A..99L, 2018MNRAS.480.3140W, 2018MNRAS.481.2901J}. For clump A, the {\tt MARS} result shows an $\mytilde100$~kpc offset with respect to the nearest BCG. At this moment, it is difficult to assess the reality of this large offset because no multiple images have been identified near the galaxy.
While substructure G is reported in some previous studies, clump F is only claimed in \citet{2018MNRAS.480.3140W}, which is also based on the free-form method.

{\bf MACSJ1149.} The large-scale cluster mass distribution is elongated in the northwest-southeast direction. The most significant mass peak (A) is well-aligned with the BCG. {\tt MARS} reveals overdense regions B and C, which show good agreement with the cluster galaxies. 
As we mentioned above, for this cluster, we set the size of virtual knots to $2\arcsec$, which is 2 times larger than in other clusters. When we reconstruct the mass model of MACSJ1149 with $1\arcsec$ virtual knots size, our result mass map shows a number of small-scale fluctuations.
It is difficult to identify many of the knots as belonging to the same sources and to determine their accurate locations because their morphological features are not sufficiently distinct and the region is severely affected by the intracluster light or bright galaxies.

\subsection{Critical Curve}\label{sec:magnification_result}
\begin{figure*}
\centering
\includegraphics[width=0.9\textwidth]{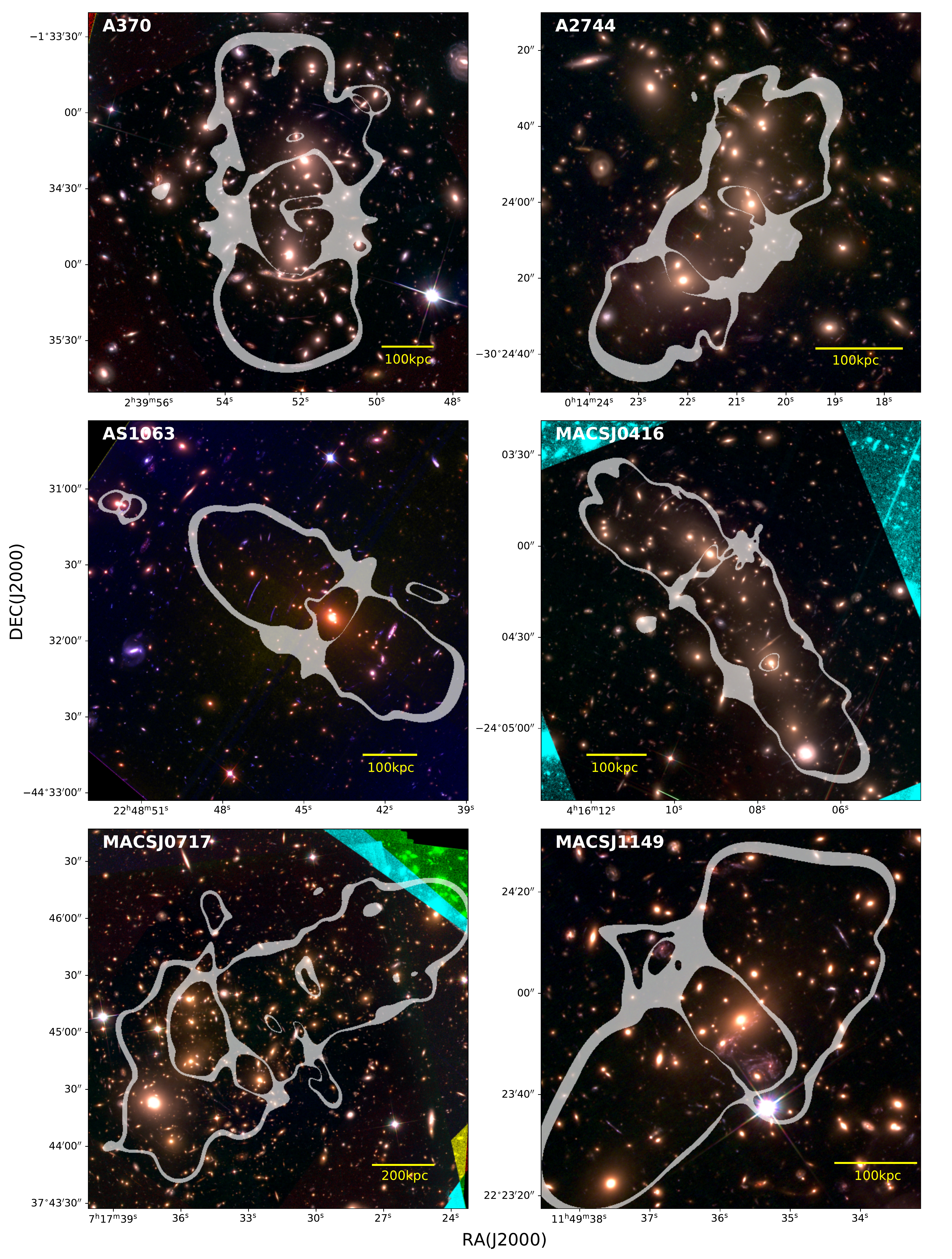} 
\caption{Critical curves of the HFF clusters. The white curves indicate the regions where the magnification is larger than 50 at the reference redshift $z_{f}=9$. The color-composite images are the same as in Figure~\ref{multiple_catalog_fig}.}
\label{result_critical_curve}
\end{figure*}

Figure~\ref{result_critical_curve} presents critical curves of the HFF clusters from our mass models. For all six clusters, the result shows both the inner and outer critical curves.
Overall, their global shapes are similar to those in the literature. 
However, significant differences are found when details are compared. This is not surprising because the critical curves trace the region where the lensing Jacobian diverges and thus a small difference in mass density can lead to a large discrepancy in magnification.

In general, critical curves from parametric models are much smoother than
free-form models. For a free-form method, {\tt MARS} produces a relatively smooth curve, thanks to its entropy-aided regularization. Compared to the parametric models, it, however, lacks details around the compact halos at the location of galaxies. This is due to the limited 
resolution shared by all free-form methods.

\section{COMPARISON WITH PREVIOUS STUDIES} \label{sec:comparison}
\subsection{Radial Profile}\label{sec:radial_profile}
\begin{figure*}
\centering
\includegraphics[width=0.88\textwidth]{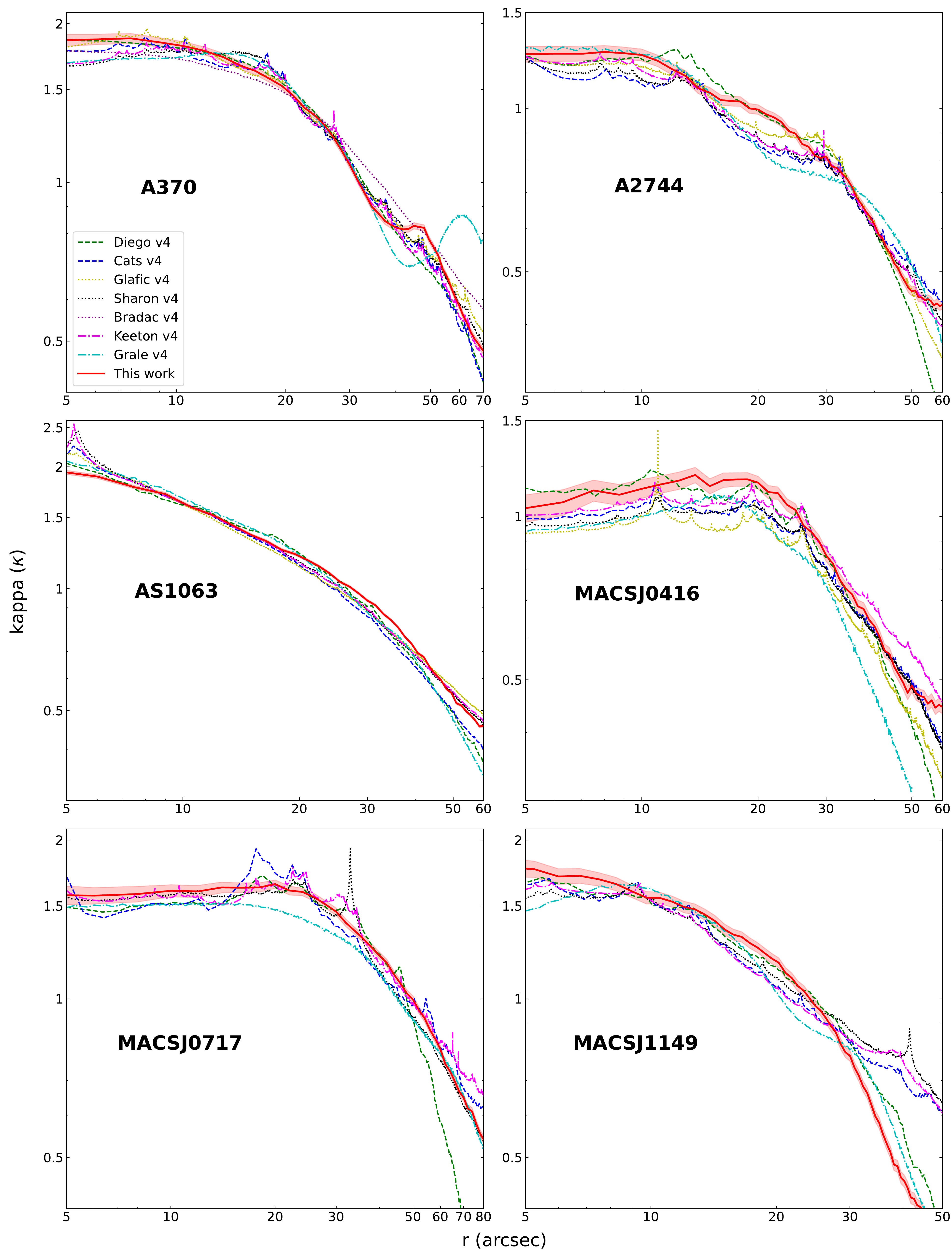} 
\caption{Radial $\kappa$ profile comparison.
We adopt the definition of the center
published in \citet{2017ApJ...837...97L}. The red solid lines represent 
our result, which is compared with those 
from Diego (green dashed lines), CATS (blue dashed lines), GLAFIC (yellow dotted lines), Sharon (black dotted lines), Keeton (pink dash-dotted lines), GRALE (cyan dash-dotted lines), and Brada{\v{c}} (brown dotted line). The red shades indicate the uncertainties.}
\label{radial_kappa_comparison}
\end{figure*}

\begin{figure*}
\centering
\includegraphics[width=0.88\textwidth]{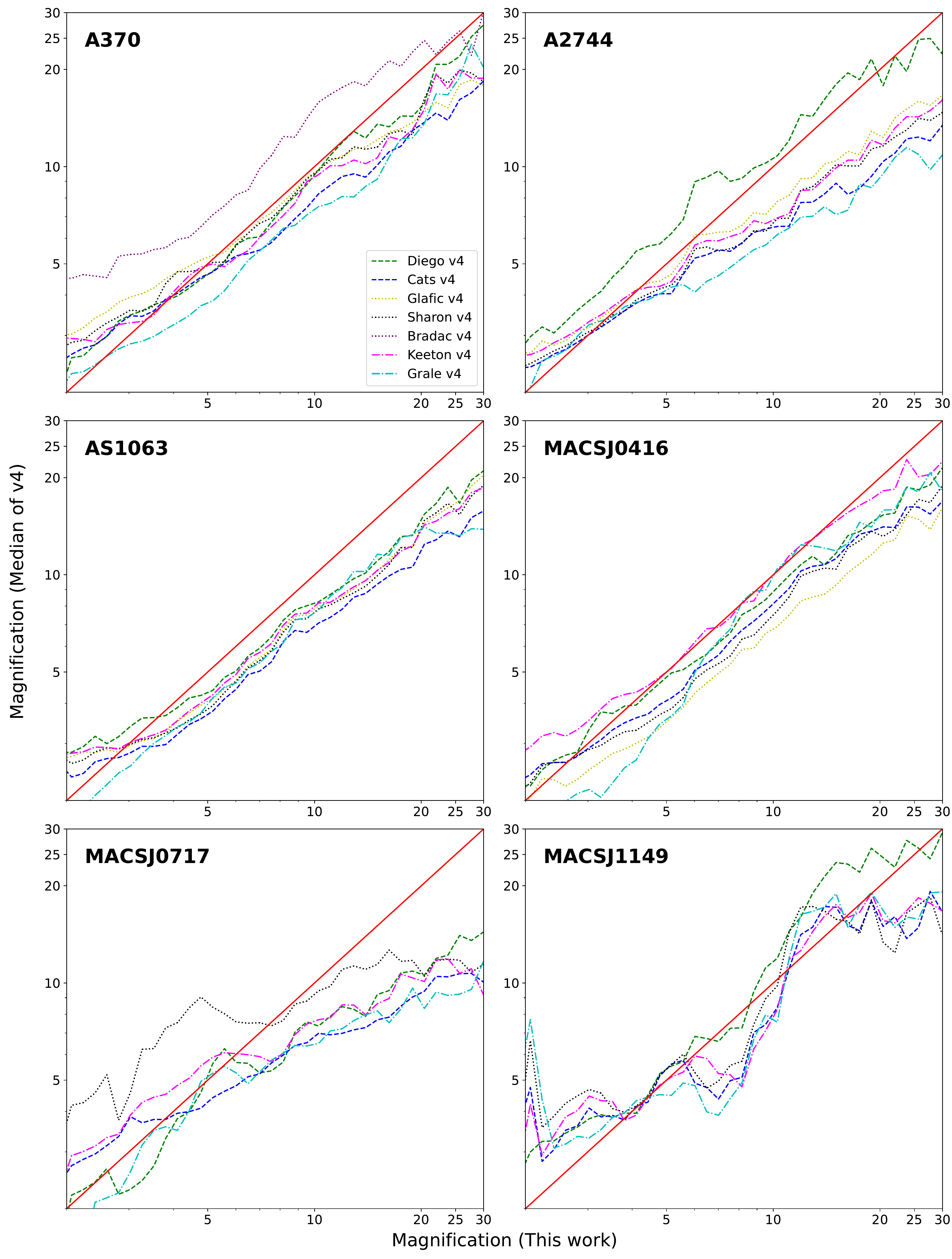} 
\caption{Magnification comparison.
We investigate the correlation of the magnification between our result and the publicly available models. 
We identify the regions in our magnification map where the value falls into the interval $\mu$ and $\mu+\delta \mu$, read the magnification values in the literature results from the same regions, and locate the median.
The red solid lines indicate the perfect 1:1 correlation case. The reference redshift for the computation of the magnification is $z=9$.
}
\label{mag_correlation_comparison}
\end{figure*}

In Figure~\ref{radial_kappa_comparison}, we compare the radial $\kappa$ profiles of the HFF clusters from different SL mass models with ours. We
adopt the cluster centers reported in \citet{2017ApJ...837...97L} when setting up radial bins.
Since $\kappa$ depends on the reference redshift, we chose $D_{ds}/D{s}=1$, which is also the convention followed by the publicly available SL models that we consider here. Uncertainties of our $\kappa$ maps are computed using the Hessian matrix described in \citet{2022ApJ...931..127C}. We clarify that the uncertainties do not include the systematic errors from the employed regularization or artifacts due to the method limitation (see \textsection2.2 in \citet{2022ApJ...931..127C} for more details).

Although we do not use identical multiple images for reconstruction, in general, our radial density profiles are roughly consistent with the results in the literature; for AS1063, all seven mass models provide highly consistent results.
In the case of A370, the radial profile from {\tt GRALE} shows a bump at $r\sim 65\arcsec$ where there are no multiple images. The authors claim that potential mass overdensities outside of the reconstruction field might cause the fictitious feature in their SL model \citep{2021MNRAS.506.6144G}. Our {\tt MARS} reconstruction, which is also based on the free-form method, does not support their claim.

One outstanding feature in parametric mass-model results is the presence of ``spikes". 
This is obviously because the parametric approaches assume that analytic halos such as the NFW profile are present at the location of the luminous galaxies. Together with the other free-form results, our mass profile from {\tt MARS} does not feature these spikes.

\subsection{Magnification}

When SL clusters are used as a cosmic telescope to observe an object in the distant
universe, one of the most critical parameters affecting the interpretation is the magnification factor, which determines the intrinsic luminosity of the object.
To compare our magnification map with those in the previous studies, we adopt the scheme used by \citet{2017MNRAS.472.3177M}; we identify the regions for each magnification interval between $\mu$ and $\mu+\delta \mu$ from our magnification map, construct the magnification distribution for the same regions from the magnification maps of other studies, and locate the median of the distribution. For a fair comparison, we only consider the region enclosed by
the convex hull defined by the multiple image positions since the lensing results
outside the region are highly affected by priors/assumptions.
In Figure~\ref{mag_correlation_comparison}, we show the results obtained in this way.

{\bf A370.} Overall, our magnifications are similar to the literature mean. At $\mu>20$, the {\tt MARS} prediction is slightly ($\mytilde20$\%) higher on average. The Diego model predicts magnification values similar to ours over the large ($1<\mu<30$) interval 

{\bf A2744.} Similar to the A370 result, our magnification is bracketed by the literature predictions. The literature mean is lower than our result in the high magnification regime ($\mytilde30$\% at $\mu\sim20$).

{\bf AS1063.} The dispersion among the literature values is small in this cluster. This is somewhat expected because their radial density profiles are similar (see Figure 5).
The {\tt MARS} prediction is higher than the literature mean by $\mytilde20$\% in the $5<\mu<20$ regime.

{\bf MACSJ0416.} Similar to the AS1063 result, the scatter among the literature values is relatively small. Since the cluster possesses the best-quality SL data set among the six HFF clusters, we believe that this is not surprising. The {\tt MARS} prediction is in good agreement with the Keeton result between $\mu=5$ and 20.

{\bf MACSJ0717.} Our magnification prediction is bracketed by the literature results at $\mu<10$, but is much larger than the literature mean at higher magnification. 

{\bf MACSJ1149.} The {\tt MARS} magnification result is bracketed by the literature results at $5<\mu<30$. Our magnification prediction is lower at $\mu<5$.

Collectively speaking, the {\tt MARS} magnification predictions tend to be slightly higher ($10-20$\%) than the literature mean in the moderate magnification regime ($\mu\sim10)$. The difference is much larger at $\mu\gtrsim20$. 
However, this behavior in the high-magnification regime is not surprising because the high-magnification region (near the critical curves) is only a small fraction of the entire field and varies significantly among methods.
Since we used our results as a reference ($x$-axis), any mismatch in the high magnification regime is likely to return a lower magnification value from the literature result for our high magnification region.
We verified this by choosing one of the literature results as the reference axis and repeating the analysis.


\section{Conclusion} \label{sec:conclusion}
We have presented new mass reconstructions of the HFF clusters using the {\tt MARS} algorithm. {\tt MARS} is the free-form SL reconstruction method that maximizes cross-entropy as a regularization. {\tt MARS} suppresses spurious fluctuations and reconstructs a smooth mass map while achieving small scatters in both the source and the image planes. We compiled $100 - 200$ self-consistent multiple images for each cluster from the literature and present the resulting mass model.
Our mass map produces small scatters lower than the literature results by an order of magnitude. The source and image plane scatters are $\mytilde0\farcs1$ and $\mytilde0\farcs02$, respectively.

In general, the mass maps are highly consistent with the distributions of the cluster galaxies, although {\tt MARS} is completely blind to their locations. 
The most remarkable case is MACSJ0416, where we use 236 multiple images that all have spectroscopic redshifts and converge them with negligible image plane scatters ($\mytilde0\farcs084$). Many group/galaxy-scale substructures are revealed and their locations are in good agreement with the member galaxies.
Our overall radial profiles are consistent with the literature results, although the differences in detail are non-negligible. 

In the current JWST era, when the upcoming observations are expected to increase the number of multiple images of many SL systems by several factors, a highly flexible, robustly regularized free-form mass
reconstruction method such as {\tt MARS} and others is required to investigate the self-consistency of the unprecedentedly large SL dataset and to provide model-independent knowledge of the underlying dark matter distributions.

This  work  is  based  on  observations  created with NASA/ESA Hubble Space Telescope and downloaded from the MiKulski Archive for Space Telescope (MAST) at the Space Telescope Science Institude (STScI).
The current research is supported by the National Research Foundation of Korea under program 2022R1A2C1003130.
We compare our results with the publicly available gravitational lensing models, which are downloadable from the MAST and produced by PIs Brada{\v{c}}, Natarajan $\&$ Kneib (CATS), Merten $\&$ Zitrin, Sharon, Williams, Keeton, Bernstein and Diego, and the GLAFIC group.
STScI is operated by the Association of Universities for Research in Astronomy, Inc. under NASA contract NAS 5-26555. 

\software{Astropy \citep{astropy2013}, Matplotlib \citep{matplotlib2007}, NumPy \citep{harris2020array}, SciPy \citep{scipy2020}}, PyTorch \citep{2019arXiv191201703P}

\appendix
\section{Multiple image catalogs}\label{multiple_image_table}

\begin{deluxetable*}{cccccc}\label{multiple_catalog_a370}
\tablecaption{Multiple image catalog of A370.}
\tablehead {
\colhead{ID} &
\colhead{R.A. (J2000)} &
\colhead{Decl. (J2000)} &
\colhead{z} & 
\colhead{Class} &
\colhead{$\Delta_{rms} (\arcsec)$}
}
\startdata
1.1	&	39.967047	&	-1.5769172	&	0.8041	&	gold	&	0.01	\\
1.2	&	39.976273	&	-1.5760558	&	0.8041	&	gold	&	0.0059	\\
1.3	&	39.968691	&	-1.5766113	&	0.8041	&	gold	&	0.0123	\\
2.1	&	39.973825	&	-1.584229	&	0.7251	&	gold	&	0.0035	\\
2.2	&	39.971003	&	-1.5850422	&	0.7251	&	gold	&	0.0156	\\
$\cdots$	&		&		&		&		&		\\
\enddata
\tablecomments{The complete multiple-image catalog is available as online supplementary material.}
\end{deluxetable*}

\section{Lens-Plane Image Reconstruction}\label{lens_plane_recon}

\begin{figure*}
\centering
\includegraphics[width=0.8\textwidth]{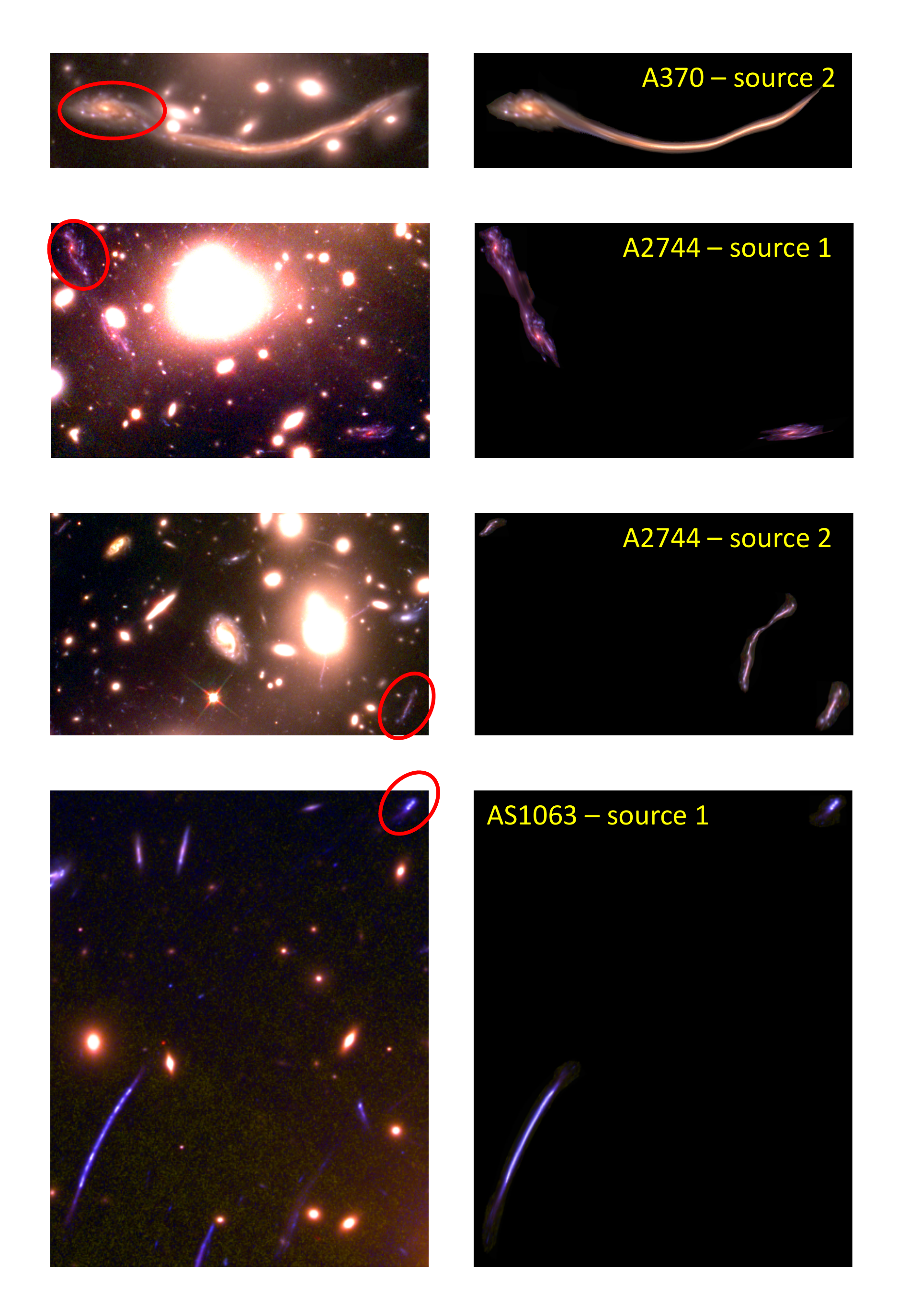} 
\caption{Lens-plane image reconstruction. Left panels show the observed multiple images. Red circles mark the images that we select for reconstruction. Right panels show the reconstructed images in the lens plane that matches the dimensions in the left panel.
For source 2 in A370 and source 1 in AS1063, we utilize only their centroids for our mass reconstruction. For sources 1 and 2 in A2744, 4 and 6 knots were used.
The overall lensed morphology and surface brightness features are well-reproduced. This high-fidelity lens-plane reconstruction would be impossible if the model became overfitted.
}
\label{multiple_recon1}
\end{figure*}

\begin{figure*}
\centering
\includegraphics[width=0.8\textwidth]{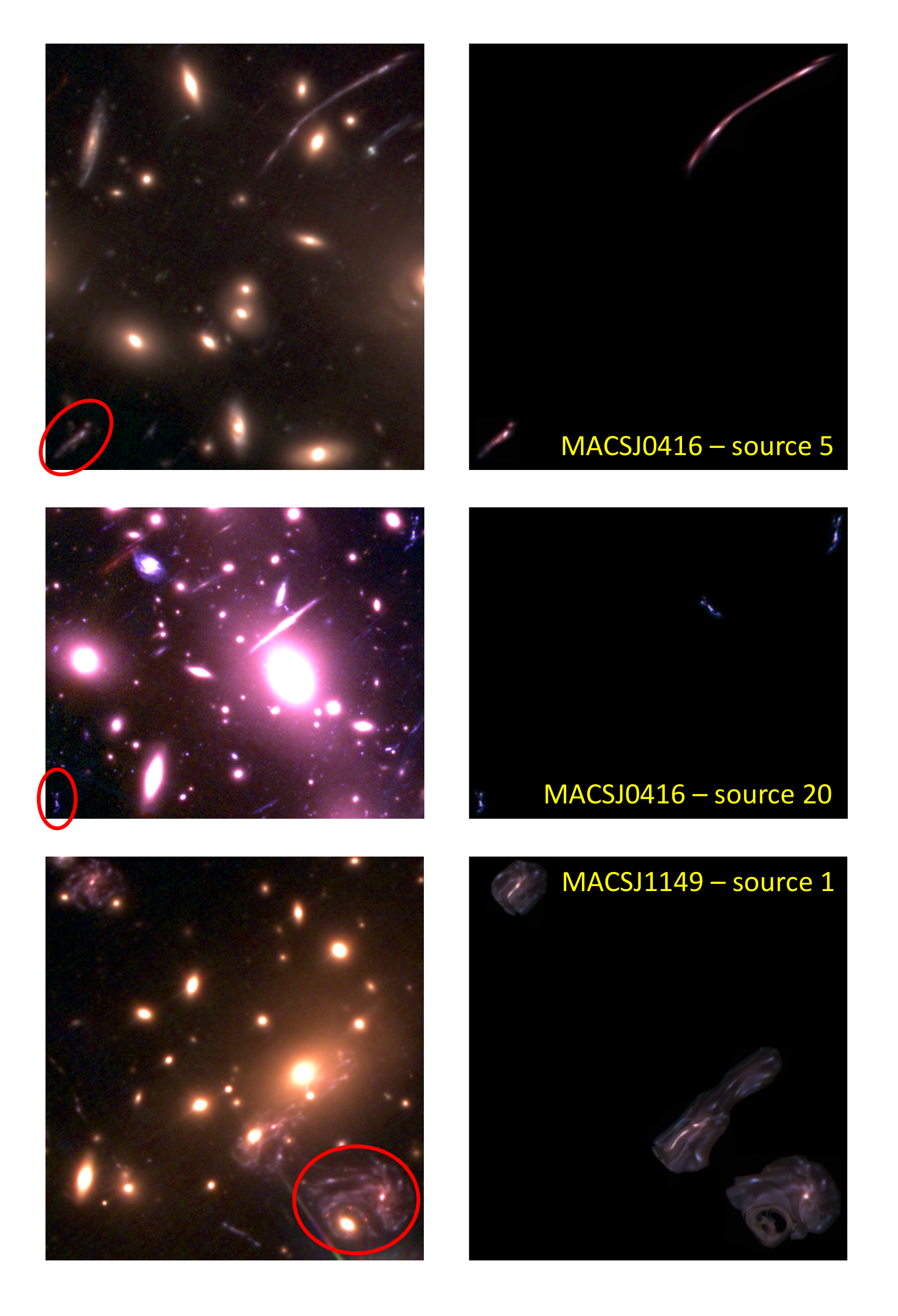} 
\caption{Lens-plane image reconstructions. Same as Figure \ref{multiple_recon1}.
For source 5 in MACSJ0416, we used 6 knots per image. For source 20 in MACSJ0416, 2, 1, and 2 knots were utilized for images 20.1, 20.2, and 20.3, respectively. In the case of source 1 in MACSJ1149, we used 30, 28, 28, 5, 6, and 1 for lensed galaxies 1.1, 1.2, 1.3, 1.4, 1.5, and 1.6, respectively.}
\label{multiple_recon2}
\end{figure*}

To demonstrate the lens-plane image reconstruction performance of {\tt MARS}, we choose the multiple image systems that are highly distorted and extended. We pick one such system per cluster from A370, AS1063, and MACSJ1149. For A2744 and MACSJ0416, two such systems per cluster are selected.
In Figure~\ref{multiple_recon1} and~\ref{multiple_recon2}, we present the lens-plane reconstruction results. The overall lensed morphology and surface brightness features are well-reproduced. 
The number of constraints (knots) used for each lensed image varies from 1 to 30.
While in general, more knots  produce better reconstructions, even a single constraint (i.e., centroid) can reasonably well reproduce the morphology and surface brightness in the lens plane.
This demonstrates that the mass model obtained with {\tt MARS} is not only optimal for minimizing the source position scatters in the lens plane, but also possesses a high-fidelity lens-plane image reconstruction capability, which would be impossible if the model significantly overfitted/underfitted the SL data.

\bibliographystyle{apj}
\bibliography{main}

\end{document}